\documentclass[manuscript]{aastex}

\usepackage{amsmath}
\usepackage{amssymb}
\newcommand{\norm}[1]{ \Vert #1 \Vert }
\newcommand{\bSeries}{\mathbf{\beta}}
\newcommand{\reals}{\ensuremath{\mathbb{R}}}
\DeclareMathOperator*{\minimize}{minimize \quad}
\DeclareMathOperator*{\subto}{subject\, to \quad}

\shorttitle{SWAP Observations of the EUV Corona}
\shortauthors{Seaton et al.}

\begin{document}


\title{SWAP Observations of the Long-Term, Large-Scale Evolution of the EUV Solar Corona}


\author{Daniel B. Seaton\altaffilmark{1}, Anik De Groof\altaffilmark{1,2}, Paul Shearer\altaffilmark{3}, David Berghmans\altaffilmark{1}, and Bogdan Nicula\altaffilmark{1}}

\affil{$^{1}$Royal Observatory of Belgium-SIDC, Avenue Circulaire 3, 1180 Brussels, Belgium}
\affil{$^{2}$ESA/European Space Astronomy Centre ESAC, P.O. Box~78, 28691 Villanueva de la Ca\~nada, Madrid, Spain}
\affil{$^{3}$Department of Mathematics, 2074 East Hall, University of Michigan, 530 Church St., Ann Arbor, MI 48109-1043}


\begin{abstract}
The Sun Watcher with Active Pixels and Image Processing (SWAP) EUV solar telescope on board the Project for On-Board Autonomy 2 (PROBA2) spacecraft has been regularly observing the solar corona in a bandpass near 17.4~nm since February~2010. With a field-of-view of 54$\times$54~arcmin, SWAP provides the widest-field images of the EUV corona available from the perspective of the Earth. By carefully processing and combining multiple SWAP images it is possible to produce low-noise composites that reveal the structure of the EUV corona to relatively large heights. A particularly important step in this processing was to remove instrumental stray light from the images by determining and deconvolving SWAP's point spread function (PSF) from the observations. In this paper we use the resulting images to conduct the first-ever study of the evolution of the large-scale structure of the corona observed in the EUV over a three-year period that includes the complete rise phase of solar cycle 24. Of particular note is the persistence over many solar rotations of bright, diffuse features composed of open magnetic field that overlie polar crown filaments and extend to large heights above the solar surface. These features appear to be related to coronal fans, which have previously been observed in white-light coronagraph images and, at low heights, in the EUV. We also discuss the evolution of the corona at different heights above the solar surface and the evolution of the corona over the course of the solar cycle by hemisphere.
\end{abstract}





\section{Introduction\label{sec:intro}}

Despite decades of observation in a wide range of wavelength bands, temperature ranges, and scales, our understanding of the connection between the high-temperature corona seen at low heights in extreme ultraviolet (EUV) and X-rays and the white-light corona most observed with coronagraphs and at solar eclipses remains tenuous. This is in no small part due to the scarcity of overlapping observations that might help better establish the relationships between low and high altitude coronal structures.

The lack of observational coverage of the extended EUV corona---and the white-light corona in the region where it overlaps EUV observations---means that very few studies of the evolution of large-scale structures in the EUV corona on long timescales have been undertaken. Those who have attempted to characterize such large-scale variations in the structure and brightness of the corona have been forced to extrapolate from available observations, both in the temporal and spatial domains, to develop a complete picture of coronal evolution.

For example, EUV observatories such as the \emph{Atmospheric Imaging Assembly} \citep[AIA,][]{2012SoPh..275...17L} on board the \emph{Solar Dynamics Observatory} (SDO) spacecraft and \emph{Extreme-Ultraviolet Imaging Telescope} \citep[EIT,][]{Delaboud95} on the \emph{Solar and Heliospheric Observatory} (SOHO) reveal the structure of the low corona and the connection between coronal features and their magnetic sources at the solar surface in high contrast, but neither EIT, with a field of view that extends to about 1.4~solar~radii along the image axes, nor AIA, whose field of view reaches about 1.3~solar~radii along the image axes and 1.4~solar~radii on the diagonal, have fields of view sufficiently large to reveal the full extent of the EUV corona. The EUV imagers of the {Sun-Earth Connection Coronal and Heliospheric Investigation} \citep[SECCHI,][]{2008SSRv..136...67H} instrument package on board the twin \emph{Solar-Terrestrial Relations Observatory} (STEREO) spacecraft have larger fields of view, extending circularly to about 1.7~solar~radii, but the quality of their observations at large heights is limited both by the image compression due to the strong telemetry constraints faced by STEREO and by the considerable instrumental stray light \citep[see][]{Shearer:2012}, which dominates at heights above 1.3~solar-radii\footnote{A comprehensive discussion of stray light in EUV observations and of the techniques used to remove it, such as those presented later in this paper, can be found in \citet{ShearerThesis} and \citet{Shearer2013}.}.

Meanwhile, the best available white-light coronagraphic observations from the Earth's perspective remain those from the \emph{Large Angle Spectroscopic Coronagraph} \citep[LASCO,][]{1995SoPh..162..357B} on board SOHO. Since the failure of LASCO's inner C1 coronagraph in 1998, the innermost observations of the corona available from LASCO are those from the C2 coronagraph, which has an inner radius of about 2~solar~radii and outer radius of about 6~solar~radii. Observations of the white-light corona from the COR1 coronagraphs \citep{2003SPIE.4853....1T}, part of the SECCHI instrument package on board STEREO, show the white-light corona between 1.3 and 4~solar~radii.

Ground-based observations from coronagraphs and eclipses provide a better view of the low corona in white light. Eclipse observations such as those by \citet{0004-637X-734-2-114} yield high-quality images that can be directly compared to EUV images \citep{2011ApJ...734..120H}, but eclipse observations are only infrequently possible. Ground-based coronagraphs can help to reveal a more complete picture of the corona, but not with the same high quality and contrast as space-based observations. Thus opportunities to produce and study high quality observation of the corona between heights of roughly 1.3~solar~radii and 2~solar~radii are limited. The absence of high quality observations of this region is no doubt a primary reason for the scarcity of studies of the evolution of the low corona at larger scales.

The \emph{Sun Watcher with Active Pixels and Image Processing} \citep[SWAP,][]{2012SoPh..tmp..217S, 2012SoPh..tmp..317H} on board the European Space Agency's \emph{Project for On-Board Autonomy 2} (PROBA2) satellite provides the widest field of view EUV images available from the viewing perspective of the Earth. SWAP is a single-channel EUV telescope with a bandpass peak near 17.4~nm and a square field-of-view approximately 54~arcmin wide, which means SWAP images reveal the EUV solar corona to a height of roughly 1.7~solar~radii along the image axes and 2.5~solar~radii on the diagonals. PROBA2 was launched on 2009~November~2, and its full-time scientific mission commenced in early February 2010, thus SWAP has observed the changing EUV corona throughout the complete rise of the solar activity cycle. SWAP's unique field of view, Earth-oriented viewing angle, and three years of continuous operation have thus yielded a unique data set that reveals the long-term evolution of a previously unexplored part of the corona. In this paper we present an analysis of these observations, which represent the first ever study of the long-term evolution of the EUV corona above heights of 1.5~solar~radii over the complete rise phase of the solar cycle.

Studies of long-term solar variability, of course, have a long history. Indeed, that the structure of the corona evolves along with the solar activity cycle has been known nearly as long as systematic observations of eclipses and sunspots have been made. It is well-known that white-light coronal steamers narrow to a small band near the solar equator during times of low activity and grow to cover nearly the entire extent of the sun during times of maximum activity \citep{1989A&AS...77...45L}. However, in recent years considerably more extensive studies of coronal variability have been made possible by full-time, space-based observations of the corona. For example, \citet{2010ApJ...710....1M} used tomographic techniques on LASCO data to reconstruct three-dimensional coronal structure over the course of nearly an entire solar cycle, perhaps the most detailed-ever look at the evolution of the solar corona at heights above 2.5~solar~radii.

The study by \citeauthor{2010ApJ...710....1M} confirms the relationship between solar activity and extended coronal structure, notably pointing out that the shift from solar minimum configuration, with large streamers confined largely to a single plasma sheet near the solar equator, to a much more complex solar maximum configuration, happens very abruptly. Extending observations of the large-scale corona above four~solar~radii to the roots of the corona near the photosphere using observations from ground-based coronagraphs, they offer some additional insight into the link between coronal sources and large-scale structure that may help shed light on this transition. Significantly, they point out that there is an association between both polar crown filaments and active regions and streamers, though streamers associated with active regions seem to be short-lived compared to streamers originating in other regions of the sun. This finding, that streamers associated with active regions are relatively short lived, stands in contrast to results reported by \citet{2002ApJ...565.1289L}, who claim that long-lived corona structures are observed to be associated with areas of magnetic activity at the solar surface.

On the other hand, in another paper, \citet{2007A&A...465L..47M} also report observations of large-scale, fan-like structures that appear to emerge from narrow, bright bundles in EUV observations of the low corona and extend to broad, diffuse structures at larger heights. These structures, which are associated with regions of open magnetic field, nonetheless rise from active regions, and appear to contain magnetic neutral sheets. SWAP observations, which we will discuss in detail below, suggest that these fan streamers can persist for many solar rotations and indeed appear to be connected to cusp-shaped structures that have long been predicted to form above active regions \citep{1967ARA&A...5..213N}. As we will see, these structures are clearly associated with the rise of solar activity and account for a significant fraction of coronal EUV emission at large heights. The ability to observe such extended structures is, in particular, a benefit of SWAP's large field of view; the lower portions of such structures can be seen in images from AIA and EIT, but to see the full extent of these features a larger field of view is required. (An example appears in overlapping AIA and SWAP observations of coronal structures associated with a white-light jet by \citealt{2013SoPh..286..143F}.)

The evolution and properties of two other coronal features have also been well-explored in white light observations, but not in the EUV: polar streamers \citep{2008ApJ...680.1532Z} and coronal pseudostreamers \citep{2007ApJ...658.1340W}. Polar streamers are, apparently, classical helmet streamers that form above the magnetic neutral line that underlies polar crown filaments. They are dipolar structures bounded by open field lines of opposite polarity separated by a current sheet. Pseudostreamers, although they superficially resemble classical streamers in appearance, are more magnetically complex than polar streamers. They contain three or more polarity regions at their base and, unlike classical streamers, the two open field regions at the outer boundary of a pseudostreamer have like polarity. 

Both types of features are associated with increasing solar activity, but neither have been extensively studied in the EUV. Pseudostreamers can often be distinguished from classical streamers by their double-lobed bases. However, in coronagraph images, where the streamer bases are not visible, making this distinction is not always easy. Thus, EUV images, such as those from SWAP, that simultaneously reveal the streamer or pseudostreamer base near the solar surface and the extended open field regions of both types of features are highy useful.

Other authors, meanwhile, have studied the relationship between the differential rotation of the sun itself and the rotation of the corona, either at very low heights using EUV images \citep[such as][]{2012SoPh..281..611L} or large heights using coronagraph images \citep[such as][]{1999SoPh..184..297L}. Both studies found that the rotation period of the corona is similar to that of the photosphere, with a rotational period of around 27~days, but that the coronal rotation was much more difficult to characterize than that of the photosphere. The exact rotation period appears to be a function both of heliospheric latitude and time, governed, apparently, by the changing magnetic structure of the corona and by disruptive events triggered by CMEs and eruptions. However, due to the complexity of the corona, none of these authors identified any simple rule that could describe the time-varying nature of coronal rotation.

Of more direct relevance to observations of the corona by SWAP was a study of coronal rotation using the \emph{Ultraviolet Coronagraph Spectrometer} \citep[UVCS][]{1995SoPh..162..313K} on board SOHO by \citet{2008ApJ...688..656G}, since they characterized the rotation of the corona at heights nearly the same as those observed uniquely by SWAP. \citeauthor{2008ApJ...688..656G} also found evidence of coronal differential rotation and temporal variation, but also reported on additional asymmetry in the rate of rotation as a function of both latitude and solar hemisphere. As we will see, SWAP confirms that the evolution appearance of the EUV corona is strongly influenced by the rotation of bright and dark regions embedded in the EUV corona. The observations we present here are not optimized for such a careful analysis of the rotation rate of the EUV corona, but we will nonetheless make some general remarks on what our observations tell us about the influence of solar---and coronal---rotation on the appearance of the EUV corona.

In this paper we present a set of specially prepared SWAP observations, corresponding to the three-year period beginning on 2~February~2010, that  reveal the structure of the far off-limb EUV corona. These observations reveal not only the evolution of the structure of the EUV corona as the solar cycle begins to increase in activity, but also a number of unique features not clearly observed before in the EUV. In Section~\ref{sec:obs} we describe the observations and the calibration steps and processing necessary to produce them. Section~\ref{sec:stray} discusses the influence of instrumental stray light on our observations as well as the blind deconvolution method used to remove stray light during processing, a critical step in producing high quality images of the extended corona. Finally, in Section~\ref{sec:disc} we present a few of the most important features in the SWAP data and discuss their implications on the role of the solar cycle in the evolution of the EUV corona.

\section{Data Processing and PSF Correction \label{sec:obs}}

PROBA2 was launched on 2009~November~1, and SWAP began capturing regular science images on 2010~February~2, near the end of the deep minimum between solar cycles~23 and 24. For this analysis we processed many SWAP images into a series of high signal-to-noise images that reveal the evolution of the EUV corona over the course of the rise of the solar cycle during a three-year period that began on 2010~February~2. In this case, we generated a single image for nearly every day in the series by carefully combining multiple nominal SWAP images.

Because the PROBA2 spacecraft platform is not stable enough to allow deep exposures to be taken without becoming blurred by spacecraft motion, SWAP images are generally limited to a maximum integration time of about 10~seconds. Additionally, since SWAP uses a passively cooled CMOS-APS detector, the detector temperature is relatively high, around 0~Celsius, so thermal noise is a major limiter of image quality for individual images. However, by carefully calibrating and combining many individual observations, we can construct a low-noise composite that reveals the very faint EUV corona to very large heights above the solar surface.

In this study we combined the images resulting from the first 120~minutes of SWAP observation for every day of the PROBA2 mission. PROBA2 generally operates with an observation cadence of roughly two~minutes, but because of spacecraft maneuvers---the spacecraft rotates 90~degrees every quarter orbit or roughly every 25~minutes---some images are lost due to motion blurring; still others may be lost due to telemetry limitations\footnote{Detailed discussions of PROBA2 spacecraft operations and of the reasons for and consequences of these maneuvers appear in \citet{2012SoPh..tmp..217S} and \citet{2013SoPh..tmp...81S}.}. Thus the two-hour observation window helped to ensure that each composite could be constructed from at least 30 images, a number sufficient to yield good results even in the noisiest (that is, the darkest) areas of SWAP's field of view.

Combining images obtained in 120~minute blocks has two other important benefits. First, because of the spacecraft rolls that occur every 25~minutes, the composites are constructed from images obtained in all four possible spacecraft orientations. This helps reduce the effects of the small amount of spatial anisotropy that remains in SWAP images after the calibration process. Second, while the observation window ensures sufficient images for good results, it is short enough that the effects of large-scale coronal evolution and solar rotation are relatively small. In two hours, structures near disk-center move only a few SWAP pixels due to solar rotation, while near the limbs these structures' motion is barely detectable. Thus, while solar rotation blurs the resulting composites a small amount near disk-center, the extended off-limb features that are the main subject of our analysis remain essentially static, helping ensure that the fine structure of the corona remains relatively sharp in the image composites.

Our procedure worked as follows: We selected all of the images obtained in the first 120~minutes of observation for every day in our study period and calibrated them using the {\sf SolarSoft IDL} routine {\sf p2sw\_prep.pro}. This software performs dark current subtraction, flat field correction, PSF deconvolution, and applies image corrections to ensure the sun is centered and rotated with its north pole up in the image frame. It is worth pointing out that of these calibration steps, PSF deconvolution is of particular importance, since SWAP images are strongly influenced by stray light at the large heights we discuss in this paper. Since this step also represents a new and significant improvement to the original SWAP calibration procedure described by \citet{2012SoPh..tmp..217S}, and is a step essential to an accurate analysis of EUV coronal brightness, we will discuss it in detail in Section~\ref{sec:stray}.

With the individual SWAP images fully calibrated, we  combined these images using median stacking---that is, we constructed a new composite image by computing the median value of every pixel in the subset of images selected for each day. By computing the median value rather than the mean we suppressed random noise in the images, rejected one-time events like cosmic-ray hits, and excluded short-duration dynamic events, yielding a set of high signal-to-noise images that show essentially only the corona's most stable features over time.

\subsection{Stray Light in SWAP Images\label{sec:stray}}

Although stray light makes a relatively small contribution to the total signal in SWAP images, the effects of stray light are most pronounced in the dark regions nearest the edge of SWAP's field-of-view that we wished to explore in this study. In tests using images obtained during lunar transits of the sun described by \citet{2012SoPh..tmp..317H}, stray light was the dominant signal source at heights above roughly 1.5~solar-radii. Furthermore, since stray light is strongly influenced by the presence of bright regions on the solar disk, the evolution of active regions on the disk can have a strong effect on the overall image brightness in dim regions at large heights. As a result, any attempt to characterize the evolution of structures high in the corona must account for---and, if possible, remove---this stray light in order to accurately assess the true brightness of extended structures in the corona.

There are several possible sources of in-field stray light in SWAP images: scattering due to non-specular reflection off of microrough mirror sources, pinholes in the front filter, and back reflection by the focal plane filter. Ray-tracing simulations presented by \citet{2012SoPh..tmp..317H} show that there are no front-filter pinholes and that there is no significant contribution to stray light from internal reflection. Thus the only significant source of stray light is non-specular reflection by the multilayer mirror coatings, which has the effect of broadening the instrument's point spread function (PSF).

Mirror surface irregularities produce a PSF with a narrow core but very broad and shallow wings. It is these wings that spread the thin layer of stray light across the image, but because they are so shallow, the Sun is the only EUV source bright enough to make their contribution noticeable. Scattering from these sources is spatially invariant above the pixel scale, so stray light contamination can be well-modeled by convolution of the unknown clean image with the PSF. Conversely, the clean image may be restored by \emph{deconvolving} the PSF from the contaminated image \citep{Starck02}, but the PSF must first be estimated using information from the images themselves.

The best information about the wings of EUV telescope PSFs comes from transit images of the Moon and the inner planets. These objects do not emit in the EUV, but EUV images of them contain scattered light from the Sun. \citet{DeForest:CoronaStrayLightTRACE:2009} were first to exploit this idea: they used images obtained during a Venus transit to fit a truncated Lorentzian model for the wings of the PSF of another EUV imager, \emph{Transition Region and Coronal Explorer} \citep[TRACE][]{1999SoPh..187..229H}. Later, the PSFs for the four filter bands of EUVI-B, the EUV imager aboard STEREO-B, were determined by \citet{Shearer:2012} using a lunar transit. These authors used a more sophisticated model of the PSF wings and determined the uncontaminated image and the PSF simultaneously, eliminating the need to guess the uncontaminated image. For SWAP we adopt a slightly simplified version of the method described by \citet{Shearer:2012}

To obtain a PSF from the SWAP lunar transit, we first defined a simple but physically reasonable PSF model. Simplicity ensures that the fit is well-constrained by the available transit data and robust to non-stray light effects such as detector nonlinearity. We then determine best-fit values for the PSF parameters by solving a (semi-)blind deconvolution problem in which the brightness of the lunar disk of the stray light corrected images is constrained to be zero.

\begin{figure}[ht!]
\centering
\includegraphics{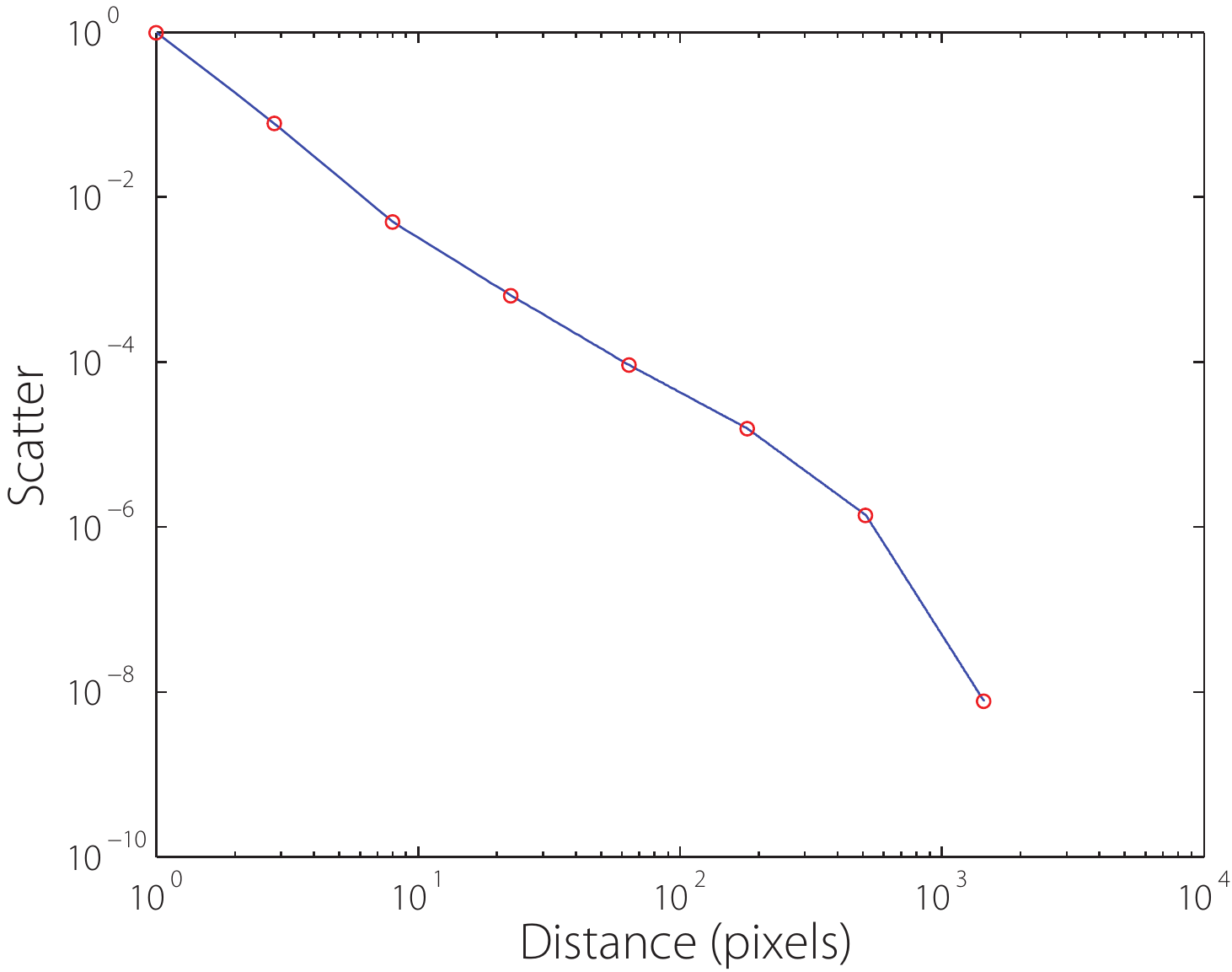} \\
\caption{A cross-section of the measured SWAP PSF showing the rate of scatter as a function of distance, in pixels, from the PSF center. The breakpoints in the profile are denoted by circles.}
\label{fig:psf_profile}
\end{figure}

We chose six images from an eclipse observed by SWAP on 2011~June~1 and obtained the PSF shown in Figures~\ref{fig:psf_profile} and \ref{fig:2d_psf}. A detailed discussion of the procedure used for modeling the PSF appears in Appendix~\ref{sec:psf_app}.

\begin{figure}[ht!]
\centering
\includegraphics{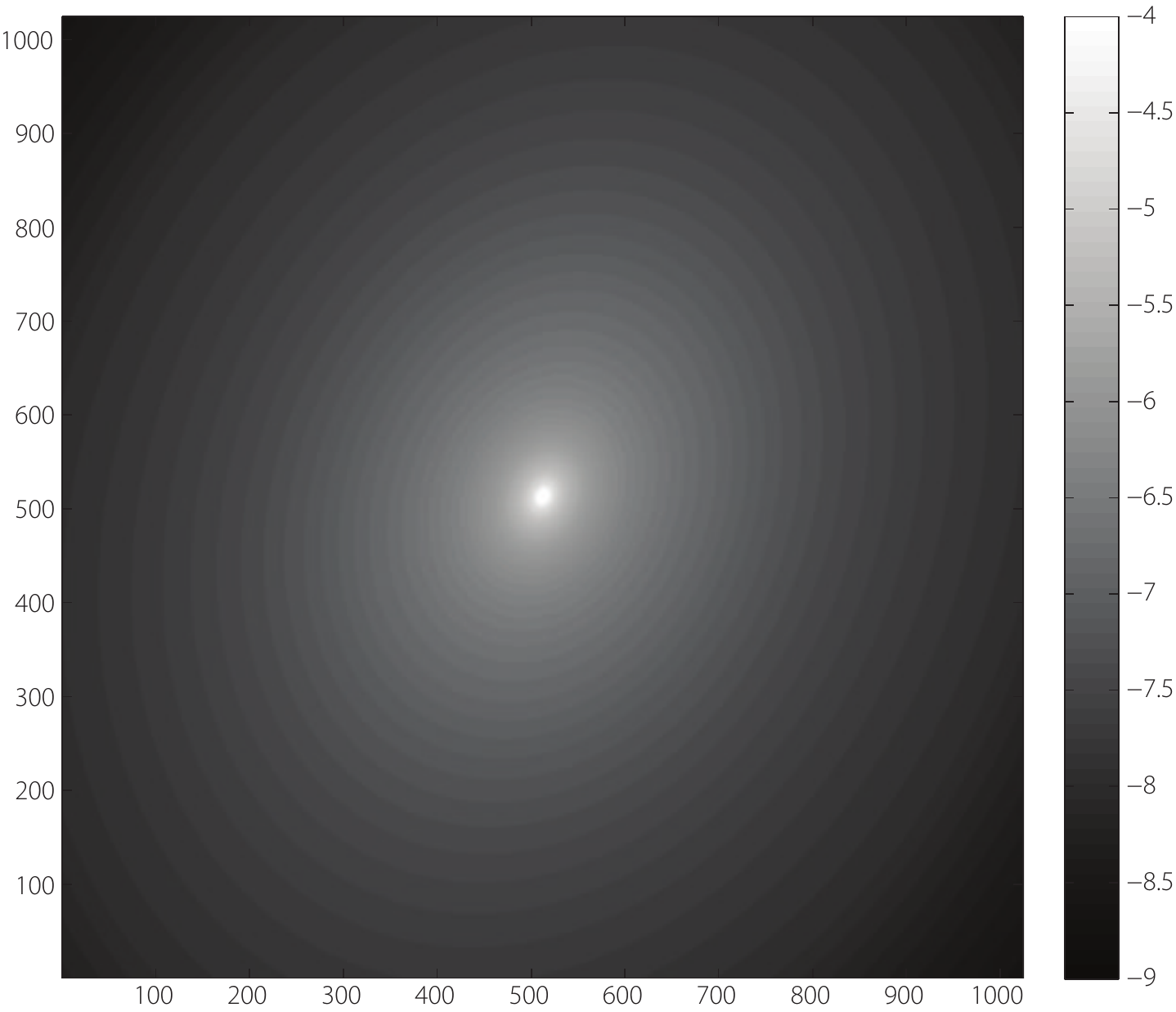} \\
\caption{A two dimensional rendering of the PSF plotted in Figure~\ref{fig:psf_profile} showing both the rate of scatter as a function of distance and the degree of anisotropy in the PSF.}
\label{fig:2d_psf}
\end{figure}

\subsection{PSF Deconvolution \& Validation\label{sec:decon}}

Using our PSF, any SWAP image $f$ can be corrected by a fast Fourier domain deconvolution. We pad both the image $f$ and the PSF $h$ to twice their size with zeros, compute the Fourier transforms $\hat{f}$ and $\hat{h}$, take the inverse Fourier transform of the quotient $\hat{f}/\hat{h}$, and then remove the padding. (As we discuss in Appendix~\ref{sec:psf_app}, there is no need for regularization because $\hat{h}$ never approaches zero.) This algorithm takes less than a second to deconvolve a SWAP image and is practically indistinguishable from the more computationally intensive conjugate gradient method used by \citet{Shearer:2012}, although the latter may be more accurate if significant signal exists near the image boundary.

\begin{figure}[ht]
\centering
\includegraphics[scale=0.5]{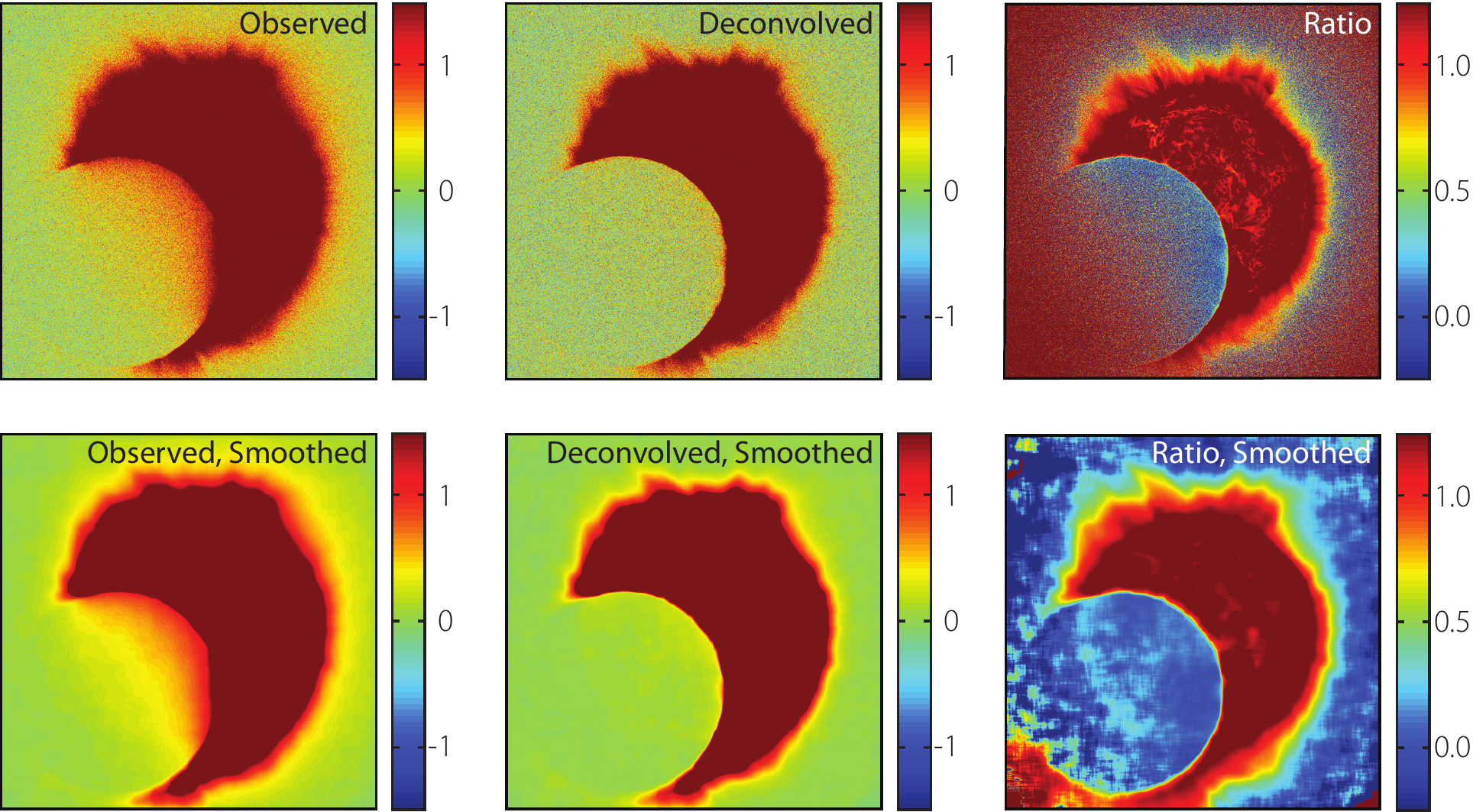}
\caption{Comparison of lunar eclipse images before stray light correction (\emph{left column}) and after (\emph{middle column}) in units of DN. The color bar has a very low upper limit and is symmetric around zero to make stray light and positivity violations easy to see.  The bottom row presents the same images smoothed with a $32 \times 32$ boxcar to reduce noise. The right column displays the ratio of deconvolved to observed (smoothed) images. The bright curved feature inside the dark moon in the ratio image in the lower right is ghost image of the solar limb, apparently the result of the low-light performance of SWAP's CMOS detector. (A complete discussion of SWAP's detector and the causes of image ghosts in SWAP images appears in \citealp{2008SoPh..249..147D}.)
}
\label{eclipse_valid}
\end{figure}

One simple validation step is to see whether the deconvolution makes sense on lunar transit images, particularly images that were not used to fit the PSF. Deconvolution with a correct PSF will make the lunar disk brightness zero and, simultaneously, will not produce any substantial violations of positivity. In Figure~\ref{eclipse_valid}, which shows a comparison of uncorrected and PSF corrected lunar transit images, we see that this is indeed the case. The Moon obscures all radiation when it passes in front of the Sun, so perfect images of the lunar disc during an eclipse should show a sharply defined and completely dark Moon superimposed on the bright sun. However, the uncorrected leftmost panels of the figure reveal broad, diffuse brightness extending far over the lunar limb --- the result of SWAP's broad PSF spreading brightness far from its origins in bright coronal structures. The corrected images, however, show a sharply defined, dark lunar limb, a confirmation that our PSF correction is performing well.

\begin{figure}[ht]
\centering
\includegraphics[scale=0.5]{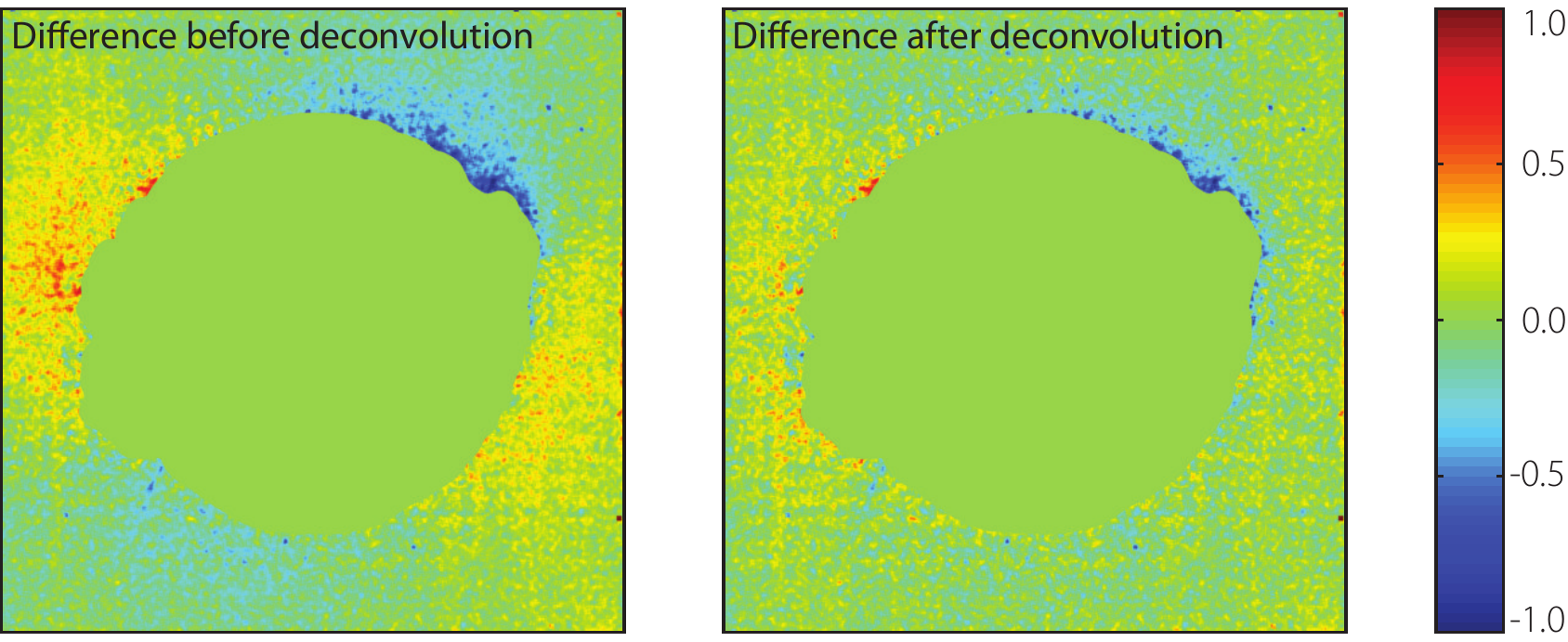}
\caption{Difference of Sun-aligned roll images before stray light correction (\emph{left}) and after (\emph{right}), units of DN. An $8 \times 8$ boxcar was applied to make the large-scale stray light distribution visible above the image noise.
}
\label{cal_roll}
\end{figure}

In another validation step similar to one used by \citet{Shearer:2012}, we independently confirmed the direction and magnitude of the anisotropy by comparing SWAP images taken at different spacecraft orientations. Since stray light is instrumental in origin, it appears to move with the solar image if the spacecraft is rotated. This phenomenon is most apparent if the difference between two Sun-aligned images is taken: such difference images reveal both a broad, bright band along one axis and a dark band along another axis in the off-limb region, which serve as an indication of the dominant direction of scatter in each image (Fig.~\ref{cal_roll}, left). Deconvolving images with the PSF greatly reduces the anisotropy along these two axes (right).

\section{Discussion\label{sec:disc}}

Calibrating the data---including the PSF deconvolution described above---and the specific techniques discussed in Section~\ref{sec:obs} yields a ``movie'' consisting of one high signal-to-noise SWAP image per day for the first three years of PROBA2's formal scientific mission. This movie constitutes a unique view of the extended EUV corona's evolution over the rise of solar cycle~24, which was at a minimum at the beginning of SWAP's nominal mission in February 2010 and had reached a relatively active state---if not a maximum---three years later. We used this movie to explore the link between increasing solar activity and the extended EUV corona and the relationship between long-lived EUV structures and long lived white-light coronal features such as those reported by \citet{2010ApJ...710....1M}.

\begin{figure}[!hp]
\centering
\includegraphics[scale=0.5]{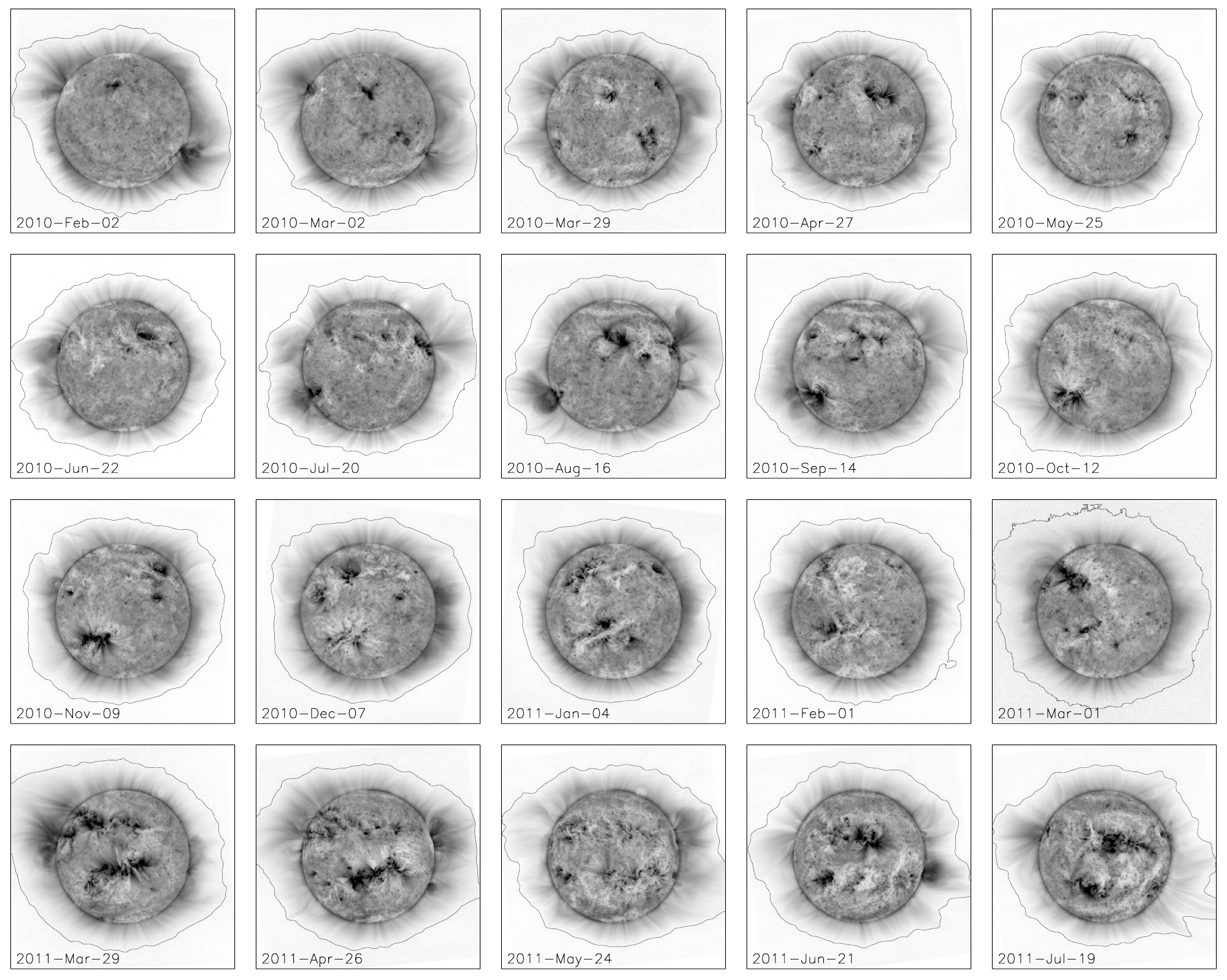}
\includegraphics[scale=0.5]{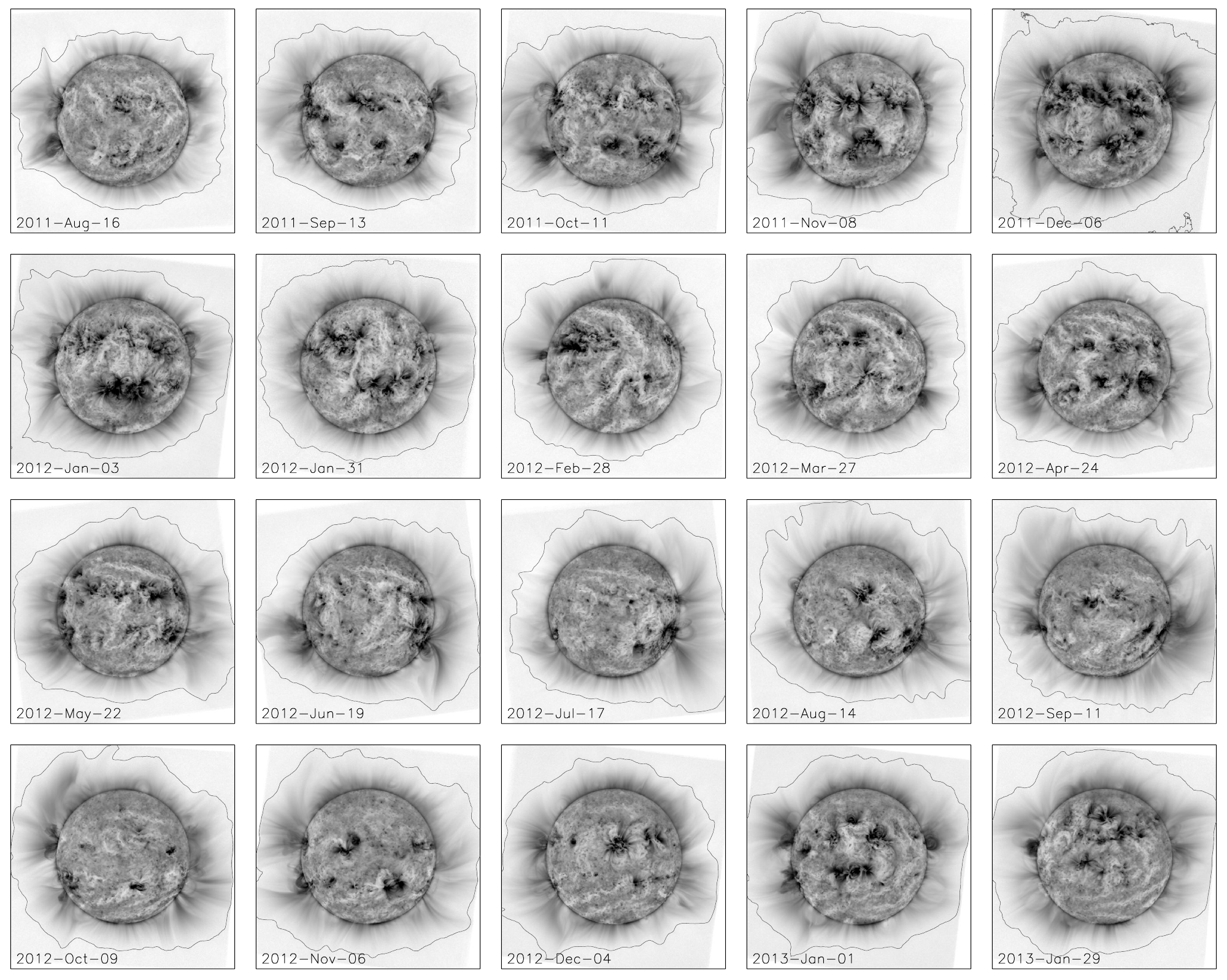}
\caption{Selected images of the corona over the course of our observation period. The contour line in each image marks the approximate location at which corona can no longer be distinguished from the image background noise. Note the considerable expansion of this boundary from near solar minimum at the beginning of the sequence to the periods of increased activity near the end of the sequence.}
\label{fig:movie_extract}
\end{figure}

Figure~\ref{fig:movie_extract} shows a series of images selected from this set, illustrating the extent and development of the corona during our study period.  (A movie of the corona's evolution corresponding to this figure is available in the on-line version of this article.)  The overplotted contours indicate the approximate location of the most distant coronal features that could be separated from the residual background noise in each image. Note that this figure---and all of the printed coronal images in this paper---is shown in inverse color scale to maximize the visibility of the faint structures that are the focus of this paper. Additionally, the images shown in the figure have been treated with a spatial filter in order to compress the substantial dynamic range of the image so it can be displayed clearly, however all of analysis described in this paper was performed using the unfiltered image sequence.

Due to spacecraft special operations and occasional telemetry gaps, it was not possible to construct an image for every day in the study period. In Figure~\ref{fig:movie_extract} and the accompanying movies, these days are simply omitted. However, for the light curve plots in the several sections that follow we have used a spline interpolation to compute approximate values for the missing days in the series. Since we have smoothed the data in these plots, this interpolation has a negligible effect on the appearance of the curves.

The transition from solar minimum to the more active conditions at the end of our study period is very clear. The number of active regions on the solar disk increases steadily during the first half of the study period and, correspondingly, the extent of the corona increases at the same time. As solar activity increases, the complexity of the structure of the corona increases as well; the EUV corona, both on-disk and off-limb, is relatively uniform in intensity and structure early in the study, but is highly heterogeneous by the end of the study, dominated by bright, extended, but localized structures.


\subsection{Periodicities in Coronal Brightness\label{sec:periodicity}}

\begin{figure}[!hp]
\centering
\includegraphics[scale=1]{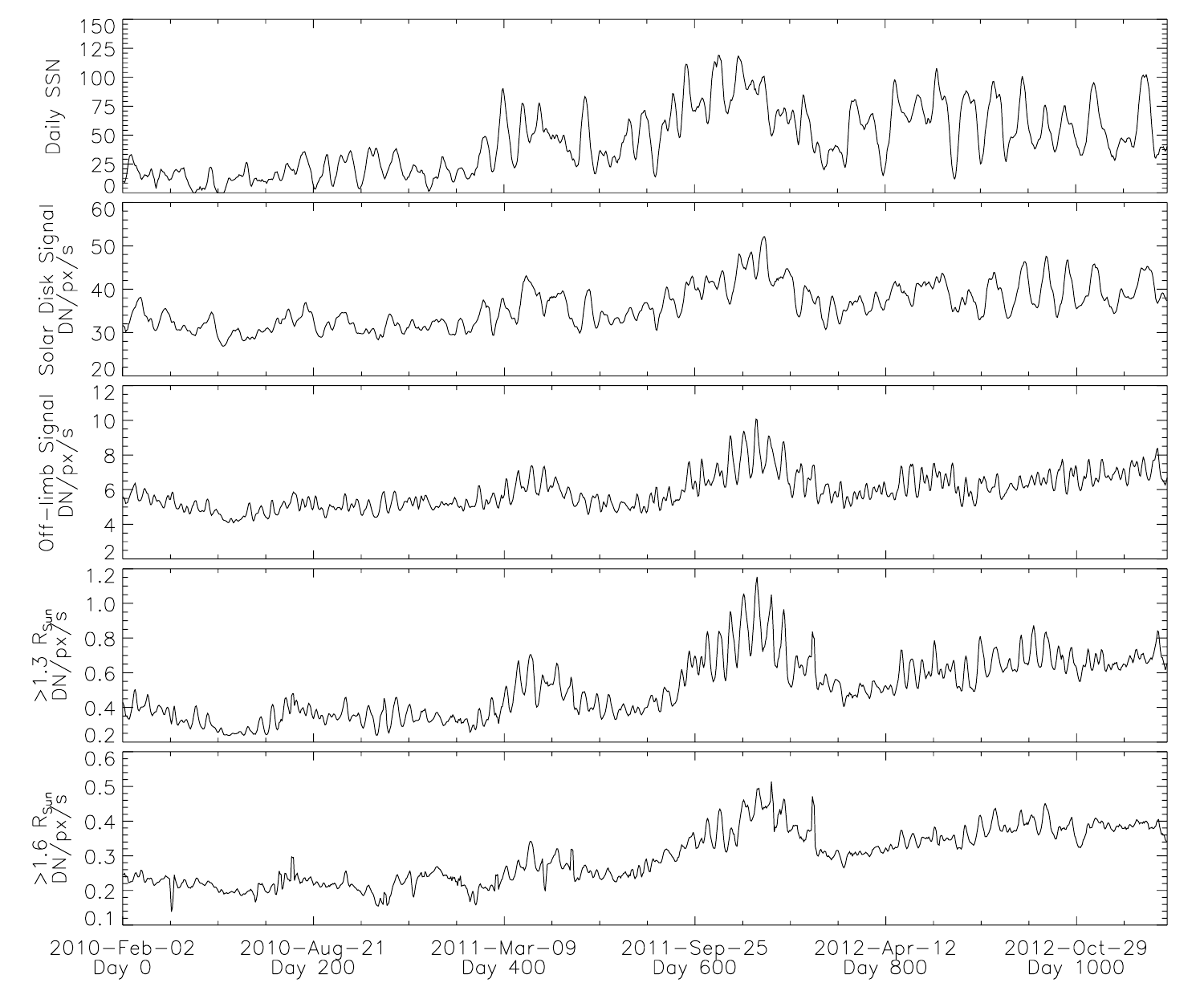}
\caption{Sunspot number and EUV signal as a function of time for the full duration of our study. The top panel shows the SSN, second panel shows the on-disk EUV signal measured by SWAP in DN/s/px, and the subsequent panels show the mean off-limb brightness above 1~solar-radius, 1.3~solar radii, and 1.6~solar radii, respectively. Strong periodicity, due to the appearance and disappearance of bright structures as a result of solar rotation is easy to see, particularly in the off-limb plots. 
}
\label{fig:lc}
\end{figure}

Perhaps the most important question is whether the extent and structure of the EUV corona varies over the course of the solar cycle in the same way that the white-light corona does. SWAP observations show that while the EUV is indeed variable over the solar cycle, evolution is characterized more by increasing extent and heterogeneity as activity increases than the white-light corona, which transitions from a narrow belt of streamers near the solar equator at minimum to a broadly extended field of streamers that originate at all latitudes at solar maximum. To better quantify the EUV corona's evolution in brightness and extent, we measured the mean brightness of each image (in data numbers, DN) above different heights and compared them to the corresponding on-disk brightness and sunspot number \citep[SSN][]{SSN2010_2013}. Figure~\ref{fig:lc} shows a summary view of our results.

Since on-disk brightness is driven largely by the presence of active regions, it is not surprising that the brightness of the disk and SSN (top two panels) bear some resemblance; the passage of strong sunspots is closely related to the appearance of bright active regions. Both the SSN and disk brightness curves exhibit, at least at certain times, strongly periodic behavior, dominated by oscillations with periods approximately as long as a Carrington rotation. Indeed, this is likely the result of particularly bright active regions rotating in and out of view repeatedly during subsequent Carrington rotations.

\begin{figure}[!hp]
\centering
\includegraphics[scale=1]{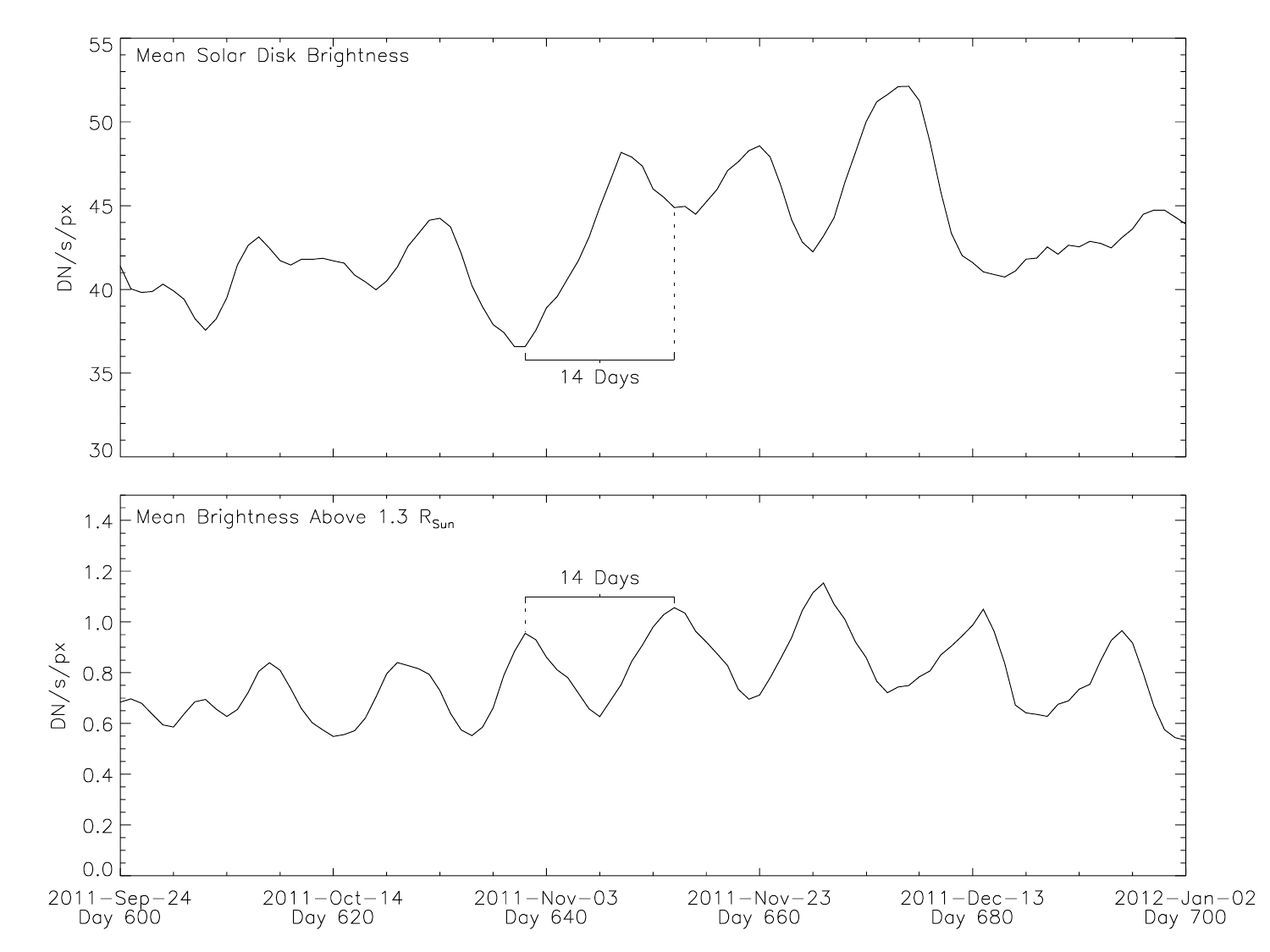}
\caption{Mean solar disk brightness (upper panel) and mean brightness above 1.3~solar~radii (lower panel) as a function of time for a 100-day period during our study. Both curves show periodicity of about 14 days, but the two curves are approximately in anti-phase. 
}
\label{fig:lc_close}
\end{figure}

More striking is the periodicity of the curves corresponding to mean signal above heights of 1, 1.3, and 1.6~solar~radii, shown in the lower three panels of Figure~\ref{fig:lc}. This periodicity is closely related to the passage of bright, extended structures from limb to limb as the sun rotates. The more sharply peaked appearance of the curves in lower panels is due to the fact that bright structures on the limb can be seen for only a few days, while bright structures on disk remain visible for roughly one week. For example, the strongly periodic behavior that occurs near the end of the study in plot of on-disk brightness can be explained by the presence of a localized bright feature low in the corona rotating onto the front and back side of the sun relative to SWAP's earthward viewpoint. Figure~\ref{fig:lc_close} illustrates the relationship between the passage of bright structures on- and off-limb.

It is apparent from Figure~\ref{fig:lc_close} that there is a relationship between the passage of bright active regions across the solar disk and bright features at large heights in the off-limb region. The periodicities of both curves are about 14 days, roughly half of one Carrington rotation, but the curves are in anti-phase. This suggests that whatever bright features are observed transiting the disk are contributing to the image brightness at large heights as they reach the solar limb. The nature of this connection, however, is not straightforward, since active regions are generally confined to low heights and do not contribute significantly to the brightness at heights above 1.3~solar~radii.

That the periodic variations are generally stronger in the off-limb region than on-disk can be explained by the fact that, while even quiet regions of the on-disk corona have some intrinsic brightness, the off-limb regions are dark except during the passage of a bright feature. As a result, the passage of locally bright features has a much more pronounced effect on the average off-limb brightness than the same feature will have on the average on-disk brightness. More surprising, however, are the long intervals during which the periodic behavior of these curves remains strongly persistent. This suggests that long-lived regions of activity remained present and relatively stationary for many solar rotations, and, indeed, several such regions can be seen rotating in and out of view in the movie that accompanies Figure~\ref{fig:movie_extract} between late August~2011 and early January~2012. This corresponds to a period of increased solar activity near the end of 2011, a period that is clearly visible in both the SSN and solar brightness plots of Figure~\ref{fig:lc} around the same period.


Such outbreaks by long-lived active regions are also the likely explanation for the several broad peaks, each a few hundred days long, in the curves shown in Figure~\ref{fig:lc}. The question of whether certain regions of activity might persist over not just many Carrington rotations, but even many solar cycles, is decades old \citep{1982SoPh...76..155B} and has been the subject of several recent studies \citep{2011ApJ...735..130L, 2013ApJ...770..149W}. Active longitudes might offer a plausible explanation for some of the surprising behavior we have observed in these plots. A more thorough analysis would be necessary to definitively settle this question---such an analysis would be an interesting application for the observations we present in this paper.

\subsection{Coronal Fans and Pseudostreamers\label{sec:fans}}

\begin{figure}[ht]
\centering
\includegraphics[scale=0.8]{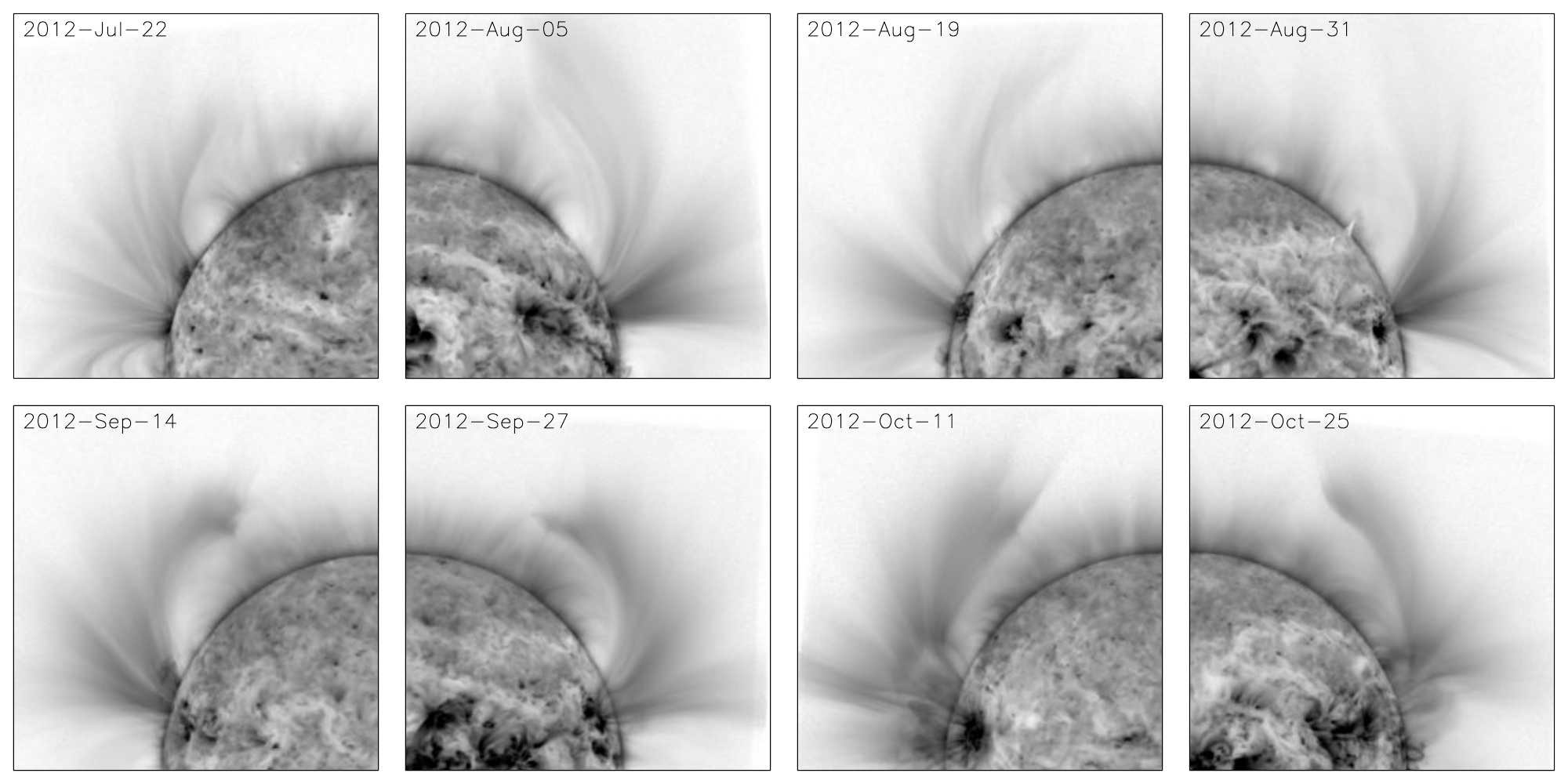}
\caption{A series of images showing the long-term evolution of a large, fan-shaped structure that persisted for at least four solar rotations in 2012. The structure was closely associated with a coronal prominence cavity, which is apparently responsible for the cusp-shaped void below the northern part of the fan structure.}
\label{fig:fans}
\end{figure}

Careful inspection of the movie accompanying Figure~\ref{fig:movie_extract} and light curves shown in Figure~\ref{fig:lc} reveals that the peaks in the brightness of the off-limb regions are often coincident with the recurrent appearance of large, fan-shaped structures that are often associated with---but not directly related to---active regions. These features can be persistent for many solar rotations and often appear to overlie cusp-shaped voids that are associated with coronal prominence cavities. Figure~\ref{fig:fans} shows an example of one such feature that persisted in the corona for at least four solar rotations at the end of our study period in 2012.

Such features have been reported in other observations, particularly observations of eclipses, for decades \citep[for a summary of such observations see][]{2002A&A...395..983K}. \citet{2007A&A...465L..47M} offered a particularly comprehensive view of these structures, presenting observations in which these so-called ``fan streamers'' could be traced from EUV images at low heights to coronagraph images that extended to at least 3~solar radii. They, too, found that these structures tend to be persistent over many days and also pointed out an association with active regions. They suggested that such structures were the result of bright, but thin sheets, viewed edge-on in the low corona, that twisted to be seen more face-on in the upper corona.

The fan rays they reported appear to be closely related to the persistent structures we see in our own observations, but it is less clear whether their hypothesis about the structure of these features can explain what we observe with SWAP.  One relatively simple test of this hypothesis is to compare the EUV view of these structures to potential-field source-surface \citep[PFSS,][]{1969SoPh....6..442S} models of coronal magnetic fields. We used the \textsf{SSWIDL} package \textsf{PFSS}, described by \citet{2003SoPh..212..165S}, to generate models of the coronal magnetic field that could be compared with our observations of these fan features.

\begin{figure}
\centering
\includegraphics[scale=1.]{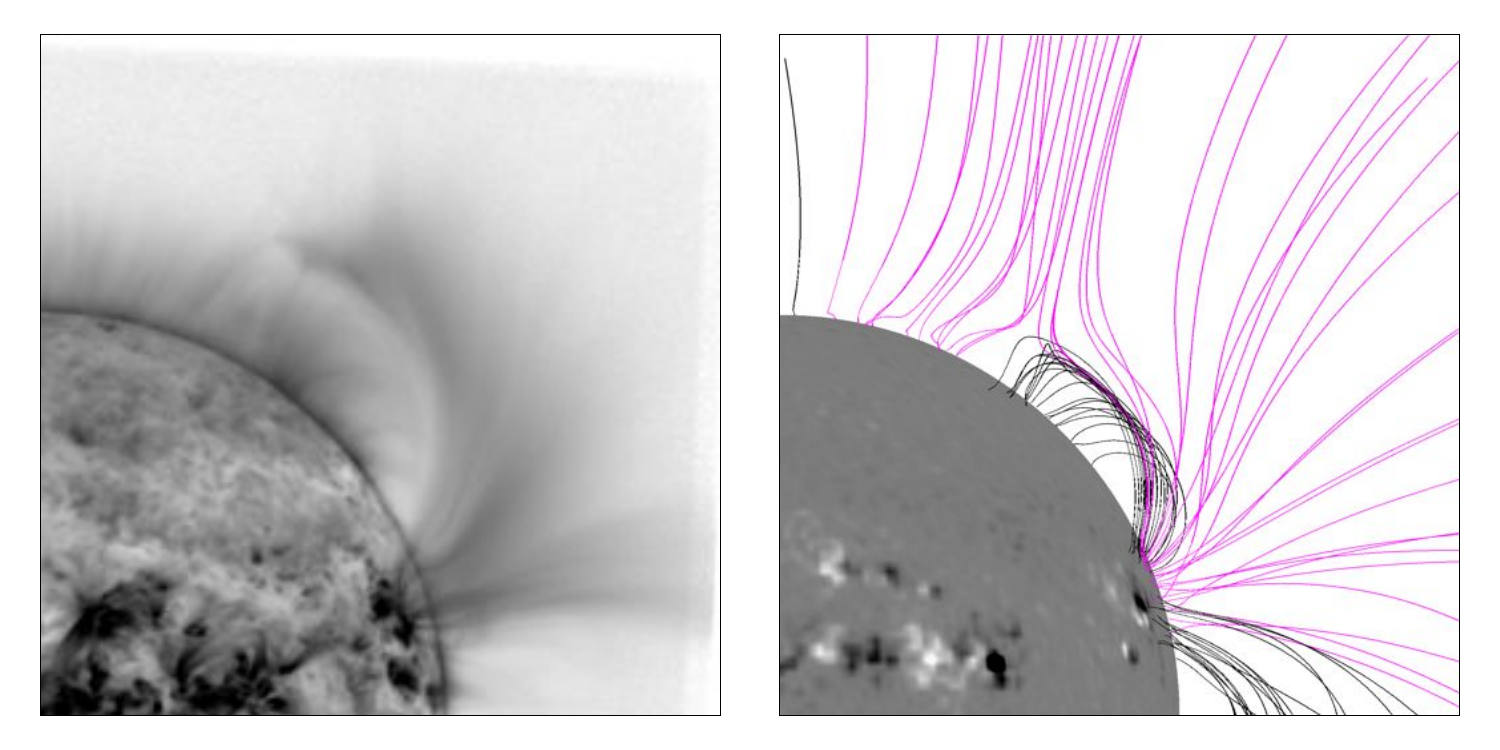}
\caption{A comparison between observations of coronal fans (left panel) and a PFSS extrapolation of the coronal magnetic field (right panel) for a fan structure observed on 27 September 2012. The purple field lines in the right panel represent open field, while the black field lines represent closed field. For clarity, we have plotted only the field-lines that originate near the limb and are thus likely to be associated with the feature we intend to study.}
\label{fig:pfss_comp}
\end{figure}

\begin{figure}
\centering
\includegraphics[scale=1.15]{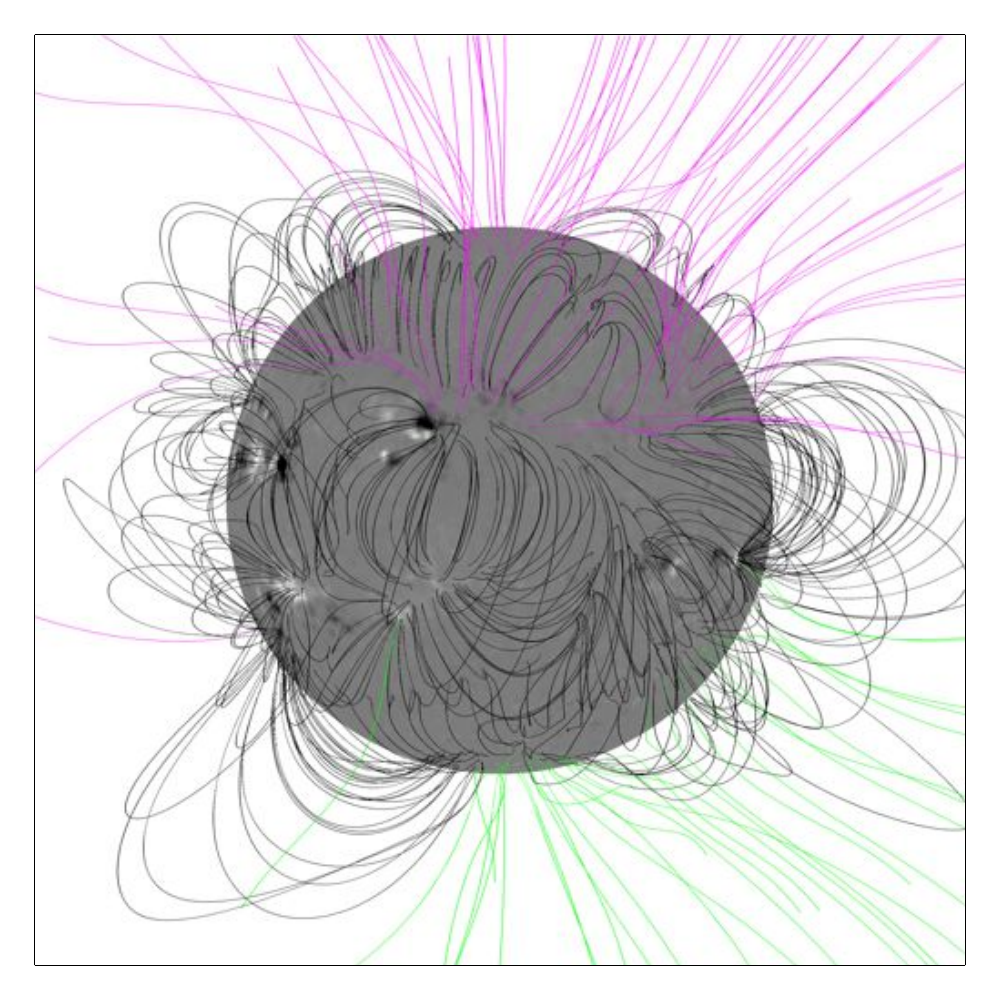}
\caption{The same potential field extrapolation as appears in \ref{fig:pfss_comp}, shown from a viewpoint rotated 90~degrees westward around the sun from the Earth perspective. This view reveals the structure of the coronal magnetic field from a perspective directly above the fan structure.}
\label{fig:pfss_lat}
\end{figure}

Figure~\ref{fig:pfss_comp} shows an example of one such comparison. The field extrapolation is not a perfect match for the observations---unsurprisingly, since much of the coronal magnetic field is not, in general, expected to be fully potential---however the PFSS model clearly captures the basic structure of what we see in the EUV observations. The purple, open field lines emerge from a narrow region at the surface of the sun and spread in much the same way the diffuse fan in the observations does. The base of this open field region is bounded to the north by a distinct region of closed field, corresponding to the void that the bright fan structure in the EUV image overlies.  If we rotate the view direction of the field extrapolation by 90 degrees so that we are able to view the fan structure from directly above (see Figure~\ref{fig:pfss_lat}) we see that both the closed field region and the open field fan structure extend a long distance around the sun.

This closed field region indeed appears to be the result of a polarity inversion line whose axis lies essentially along the line of sight in the SWAP image, while the open field region likewise extends the full length inversion line, running all along its southern boundary.  In some cases such structures are associated with filaments that run roughly along the polarity inversion, although for the specific case for which we studied the PFSS model fields there was little evidence of a filament in the observations.

In some sense this result confirms the observation by \citeauthor{2007A&A...465L..47M} that these structures are sheet-like: the fan structures observed by SWAP are indeed extended along a particularly deep line of sight. On the other hand, there is little evidence in our observations to support the hypothesis that these structures are twisted sheets, our observations suggest that they are broad and diffuse along their entire axis, with sharply defined edges that result from the boundaries set by the closed field that surrounds them.

Why are these features bright in SWAP while nearby closed field structures are more or less invisible? The answer appears to be that the closed loop system that underlies the cusp feature is simply too hot to be seen in SWAP's 17.4~nm passband, which is most sensitive to plasma just under 1~MK in temperature. SDO observations of the same feature in passbands associated with higher temperatures, 19.3 and 21.1~nm, for example, show broad, diffuse emission throughout the whole region, suggesting that the closed loops are generating emission in hotter spectral lines such as Fe~\textsc{xii} and Fe~\textsc{xiv} observed in those passbands. 

The sharp boundary that separates the bright feature in SWAP from the void below likely corresponds to the division between loops that close over the top of the polarity inversion and open field that extends to large heights. The reason for the temperature discontinuity between these two regions, however, remains an open question. Perhaps the closed-field region is heated by the dissipation of waves that are free to escape into the outer corona in the overlying open field region, resulting in a cooler average temperature in the overlying, open field region. Whatever the reason, it is clear that, as long as the magnetic connectivity between the two regions does not change, heat exchange between the upper and lower regions is not possible, which would at least  explain the persistence of this discontinuity, if not its origins.

Another important question is the extent to which these features are related to the polar streamers discussed by \citet{2008ApJ...680.1532Z} and the pseudostreamers discussed by \citet{2007ApJ...658.1340W}. In many cases, such as the one shown in Figure~\ref{fig:pfss_comp}, there is some evidence that the fan structure is associated with a pseudostreamer. The underlying void has a clearly discernible double-lobed structure consistent with that of a pseudostreamer base, and the PFSS extrapolation suggests that it is, indeed, a meeting of field lines of like polarity that is responsible for the cusp-shaped feature at the northernmost end of the bright fan. 

A similar, although less prominent, cusp-shaped feature was associated with a complex of filaments that erupted in several stages on 2010~August~1. Although this feature was near the central meridian during the eruption, it became clearly visible in SWAP images once this region rotated to the solar limb, about a week after the eruption. There is compelling evidence these eruptions took place inside of a pseudostreamer \citep{2012ApJ...759...70T}, which further suggests that these structures can be associated with pseudostreamers.

In other cases, however, it is less obvious that the fan structures observed by SWAP overlie a true pseudostreamer. For example, in several of the images of the upper panels of Figure~\ref{fig:fans}, it appears the fan originates slightly to the south of the more southward of a pair of prominence cavities. These cavities may indeed constitute the two lobes at the bottom of a pseudostreamer, but it also appears that the two cavities may be separated by a region of open field, something that is particularly apparent in the upper right panel of the figure, labeled 2012-Aug-31. If that is indeed the case, then instead of overlying a pseudostreamer, the fan structure on that particular day would have been associated with the more southerly of two classical polar streamers.

Whatever the case may be, the bright fan structures are clearly more localized than either streamers or pseudostreamers generally are. The fan structures are visible near the solar limb for only a day or two, while streamers and pseudostreamers frequently extend a large distance around the sun and can remain on the limb for days or even weeks. This highlights another important factor in the generation of these fan structures: they are almost always associated with a long-lived, localized region of activity near the edges of the closed field region the fan overlies. It appears that these small regions of activity are responsible for the brightenings that make these open-field structures so clearly visible in SWAP images in the first place. Given this, it is entirely possible that these structures can be mutually related to coronal fans, polar streamers, and pseudostreamers, depending on the local conditions where they have formed.

In both the case presented in Figure~\ref{fig:fans} and other similar cases visible in Figure~\ref{fig:movie_extract} (and the movie that accompanies it), the precise nature of the fans and the structures they overlie is not easy to determine using only SWAP data. We prefer not to speculate on the precise nature of these structures; however, it is clear that SWAP images can provide useful evidence for the interpretation of large, extended coronal features occuring near the poles, especially when used in conjunction with observations from other instruments.

Some particularly useful tools for studying these features might be provided by STEREO. STEREO's EUV images have a similarly sized field-of-view to that of SWAP, but in four different wavebands. Additionally, coronagraph images from STEREO overlap fully with the corresponding EUV images, making it possible to track connected structures from their origins at the solar surface to heights as large as 15~solar~radii. However, stray light in EUV images from STEREO is a more pervasive problem than stray light in SWAP images, and it is harder to remove it completely from the images, particularly in regions far from the solar disc \citep{Shearer2013}. Additionally, the image compression used by STEREO can alter the intensities of faint structures at large scales such as those we consider here. On the other hand, STEREO does return a few losslessly compressed images every day, so such a comparison might nonetheless be possible.

\subsection{Coronal Evolution by Hemisphere\label{sec:ns}}

\begin{figure}
\centering
\includegraphics[scale=1]{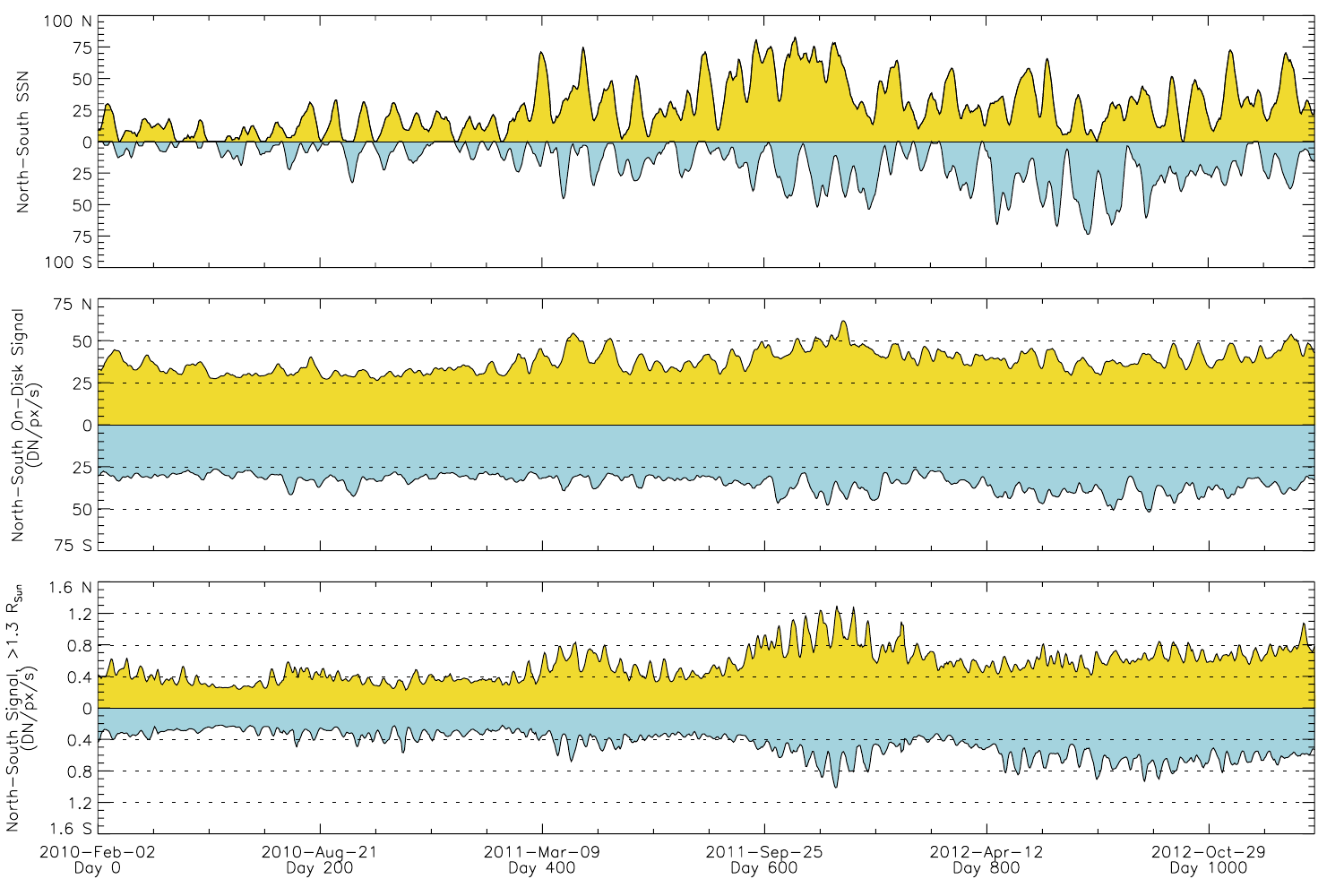}
\caption{North-South division of sunspot number (top), mean on-disk brightness (middle) and off-disk brightness (bottom).}
\label{fig:ns_comp}
\end{figure}

We also investigated whether there was any relationship between the brightness of the extended corona and the asymmetry of solar activity and sunspot numbers between the two solar hemispheres. Recent studies by \citet{2011SoPh..274..251A} and \citet{2013SoPh..282..249T}, for example, have suggested that the northern hemisphere is leading the southern hemisphere in activity, with a peak that occurred in 2011. We therefore compared the hemispheric sunspot number to the coronal brightness on-disk and off-limb for the full duration of our study period. Figure~\ref{fig:ns_comp} shows a plot of these three values, split by hemisphere and corrected for the changing solar $B_0$ angle, which varies over the course of the year due to the tilt of the solar equatorial plane with respect to the ecliptic.

It is clear from Figure~\ref{fig:ns_comp} that the northern hemisphere did experience a period of increased sunspot activity during late 2011 and early 2012. Activity in the southern hemisphere also increased during this period, but not as strongly as it did in the north. Instead, activity appears to increase in the south only towards the end of our study period in late 2012. It is also clear from the figure---and from Figure~\ref{fig:movie_extract} and its associated movie---that the mean on-disk brightness in the northern hemisphere is noticeably increased during this period of increased activity. At the same time, the brightness of the southern hemisphere, although it experiences a few short increases, is not significantly changed from the brightness detected immediately before and after the period of increased activity. At large heights, as we see in the bottom panel of Figure~\ref{fig:ns_comp} this effect is even more pronounced. The mean brightness at heights above 1.3~solar radii increases in both hemispheres, but much more significantly in the north. It is clear that EUV brightness, particularly at large heights, is related to the rate of sunspot activity.

This correlation is, perhaps, unsurprising, especially in the context of the discussion in section~\ref{sec:fans}. The fan structures observed in SWAP data are the single largest source of brightness at heights above 1.3~solar radii, and these fans are closely associated with the appearance of active regions at lower heights. Active regions, in turn, are associated with the appearance of sunspots, and thus any increase in the SSN is likely to be associated with an increase in the formation of the bright, open field structures that are observed as fans.

Nonetheless, the north-south asymmetry is considerably less clear in both the on-disk and off-limb EUV observations than it is in the SSN data. One reason for this is that, while sunspots are confined to the solar surface and, as a result, can be localized to one hemisphere or the other, the bright extended structures SWAP observes at large heights are not necessarily confined to the hemisphere from which they originate. Structures with their footpoints in one hemisphere may bend or spread so that at large heights they may contribute to the brightness of the opposite hemisphere.

\section{Conclusion\label{sec:con}}

The region of the corona that lies between the edges of the AIA field-of-view at about 1.3~solar~radii and the edge of LASCO's C2 occulting disk at about 2~solar~radii is, perhaps, the least explored region of the solar atmosphere. Prior to STEREO's launch in 2006, EUV images of this region were only rarely available.  Meanwhile, truly high quality white-light observations of this region, free of stray light and other image artifacts, were often only available at total solar eclipses; this has been especially so since the failure of LASCO's innermost C1 coronagraph. Since the beginning of PROBA2's scientific mission in 2010, however, SWAP has produced continuous EUV observations that fill this gap, the first ever complete set of EUV observations of this region taken from the perspective of the Earth.

These observations prove that the rise of the solar cycle observed at the photosphere and in the low corona is closely correlated with the growth of the complexity and extent of the EUV corona at large heights. A particular surprise, however, was that rising activity in the low corona is linked to rising activity at larger heights not through a simple extension of the growth of active regions, but through the development of long-lived, extended structures associated with open field overlying prominences near the poles and at the boundaries of active regions.

Several important questions about the nature of the corona in the poorly explored region between the low corona seen by most EUV imagers and the extended white light corona seen by coronagraphs  remain to be answered, however. Perhaps the most important is what exactly is the source of brightness in coronal structures at large heights. While there seems to be wide agreement that EUV emission low in the corona is driven by collisions between ions and electrons, there are indications that resonant scattering is the dominant source of emission at large heights \citep{2000ApJ...531.1121S}. The question, however, has not been settled \citep[see, for example,][]{DeForest:CoronaStrayLightTRACE:2009}. The SWAP observations presented here may be useful in resolving this long-standing debate.

Another question is exactly how the structures we observe in the EUV are related to the polar streamers and pseudostreamers seen in white-light observations from coronagraphs and eclipses \citep{2008ApJ...680.1532Z}. An important step in answering this question is the development of a better understanding of the three-dimensional structure of these features. At the time of this writing, efforts are underway apply the rotational tomography techniques discussed by \citet{2008SoPh..248..409B} to the observations discussed in this paper, and related data sets from SWAP, in order to answer this question. These tomographic techniques require some assumptions about the isotropy of emissivity of coronal plasma that, in turn, depend on the emission mechanism \citep{2013SoPh..283..227B}, so the two problems discussed here are closely intertwined.

While we believe it clear that SWAP is a unique, useful tool for the analysis of the large scale EUV corona, it seems equally clear the observations we have presented here raise as many questions as they answer, many of which are beyond the scope of this introductory study. It is our hope that other authors will find these and other observations from SWAP useful in their own studies. To help facilitate their use we have published all of the tools necessary to reproduce them in SWAP's \textsf{SSWIDL} software package. We have also added the PSF deconvolution procedure discussed here as an optional feature in SWAP's calibration software, \textsf{p2sw\_prep.pro}, which is also available in \textsf{SSWIDL}.

\acknowledgments

SWAP is a project of the Centre Spatial de Li\`ege and the Royal Observatory of Belgium funded by the Belgian Federal Science Policy Office (BELSPO). DBS acknowledges support from the Belgian Federal Science Policy Office through the ESA-PRODEX program, grant no. 4000103240. PS's contributions to this project were made possible by a PROBA2 Guest Investigator grant funded by ESA. We thank Laurel Rachmeler for helpful discussions about this manuscript and the anonymous referee for valuable and constructive criticism that improved both the content and clarity of this paper.



{\it Facilities:} \facility{PROBA2}.

\appendix

\section{The PSF Model\label{sec:psf_app}}

We model SWAP's PSF $h$ using a slightly modified version of the microroughness PSF model first presented by \citet{Shearer:2012}. We assume that a fraction $\alpha$ of light is not scattered, giving the PSF a single-pixel core with mass $\alpha$. The remaining fraction $1 - \alpha$ is scattered according to a power law with a decay rate $\beta$, which depends on the distance $r$ from the PSF origin. In order to model any anisotropy in the PSF, we stretch the isotropic PSF profile in one direction, transforming its circular level sets into ellipses. The resulting PSF model accommodates anisotropy and irregular wing decay using only a few parameters. 

We use this model because it is both mathematically convenient and a physically reasonable model of microrough mirror scatter. For microrough mirror PSFs, the wings are described by the power spectral density (PSD) of the mirror surface height function, and this PSD has been directly measured for other instruments \citep{Galarce:EUVModeling:2010}. Log-log plots of the measured PSD versus spatial frequency are roughly piecewise linear, and making this assumption explicitly leads directly to our proposed model. Deconvolution with a PSF from this model is especially easy because when $\alpha > 1/2$, the matrix that performs the convolution operation is diagonally dominant and hence stably invertible. 

To represent the variable exponent power law mathematically, we define a series of logarithmically spaced breakpoints $1 = r_0 < r_1 < r_2 < \ldots < r_b = r_{max}$, where $r_{max}$ is the length of the image diagonal in pixels, and define $\beta_i$ the decay exponent on the interval $[r_{i-1},r_i)$. For SWAP, the number of intervals $b$ was set to $7$, close to the value reported by \citet{Shearer:2012} in their analysis of the PSF of EUVI. With the PSF model defined, the initial isotropic PSF profile $p_{\alpha,\bSeries}(r)$ can be written
\begin{equation}
	p_{\alpha,\bSeries}(r) 	= 
	\begin{cases} 
		\alpha &\text{if $r = 0$} \\
		(1-\alpha) c_i r^{-\beta_i}, \quad &\text{if $r \in [r_{i-1},r_i)$}
	\end{cases}, \\
\end{equation}
where the $c_i$ are determined by the condition that the pieces must link together continuously. Note that if a logarithm is applied to both sides, then $\log p$ is a continuous piecewise linear function of $\log r$, so the profile can be generated on a computer by fitting a spline in log-log space and applying an exponential. The isotropic PSF generated from this profile can be written as $p_{\alpha,\bSeries}(\norm{x}_2)$, where $x \in \reals^2$.

To stretch the isotropic PSF, we define $s$ to be the stretch factor and $\theta$ the CCW angle in radians between the stretch axis and the horizontal. We apply a horizontal stretch by a factor of $s$ and a rotation by $\theta$ radians to the graph of $p_{\alpha,\bSeries}(\norm{x}_2)$. This graph transformation can be implemented by applying its inverse directly to $x$:
\begin{align} \label{hModel}
	h_\varphi(x) 			&= C \cdot p_{\alpha,\bSeries} (\norm{M_{s,\theta}^{-1} \cdot x}_2), \quad \text{where} \\
	M_{s,\theta} &= 
	\begin{bmatrix}
		\cos(\theta) & \sin(-\theta) \\
		\sin(\theta)  & \cos(\theta)
	\end{bmatrix}
	\begin{bmatrix}
		s 	& 0 \\
		0 	& 1
	\end{bmatrix}
.
\end{align}
Here $C$ is the normalizing constant defined to make the PSF sum to unity.

To determine the PSF from $N$ lunar transit images $f_1,\ldots,f_N$, we must solve for both the PSF parameters $\varphi$ and the clean lunar transit images $u_1,\ldots,u_N$ simultaneously. We require that each $u_i$ must be zero on the set of lunar disk pixels $Z_i$, and the PSF $h_\varphi$ and the $u_i$ must convolve together to form the $f_i$. This gives a large constrained nonlinear system of equations, which we will solve in the least squares sense:
\begin{equation}
\begin{split}
	f_1 	&= h_\varphi * u_1, \quad u_1(Z_1) = 0 \\
	f_2 	&= h_\varphi * u_2, \quad u_2(Z_2) = 0  \\
		& \cdots \\
	f_N	&= h_\varphi * u_N, \quad u_N(Z_N) = 0,
\end{split}
\end{equation}
where
\begin{equation}
	(u * v)(x) = \sum_{x' \in I} u(x - x') v(x')
\end{equation}
denotes the convolution of two arrays $u$ and $v$ over the index set $I$ of the $1024 \times 1024$ pixel array of CMOS detectors in SWAP. (We use zero boundary conditions, setting $u(x-x') = 0$ at pixels $x-x'~\notin~I$.) The least squares optimization problem may be written as
\begin{equation} \label{uPhiOptEqnA}
\begin{split} 
	\minimize_{\varphi = (\bSeries,\alpha,s,\theta), \, \, \{u_i\} \, \, \, \, \, \, \,} 	
	&\sum_{i=1}^N \norm{ h_\varphi * u_i - f_i }^2 \\				
	\subto \, \, \, \, \, \,
	&u_i(Z_i) = 0 \, \, \text{for all $i$,} \\
	&\bSeries \geq 0, \, \, 0 \leq \alpha \leq 1.
\end{split}
\end{equation}
Since each $u_i$ is a $1024 \times 1024$ image, this problem has over $N$ million free variables and is difficult to solve using general-purpose numerical methods. However, if $\varphi$ is set to some fixed value and the zero variables $u_i(Z_i)$ are eliminated, it decomposes into $N$ separate unconstrained linear least squares problems, each of which is easily solved by a few iterations of the conjugate gradient method. If we write the solution to each linear problem as $(u_i)_{\varphi}$, the problem reduces to
\begin{equation} \label{uPhiOptEqnB}
\begin{split} 
	&\minimize_{ \varphi = (\bSeries,\alpha,s,\theta) } 	
		\sum_{i=1}^N \norm{ h_\varphi * (u_i)_\varphi - f_i }^2 \\				
	&\subto  \bSeries \geq 0, \, \, 0 \leq \alpha \leq 1,
\end{split}
\end{equation}
which has $b+3 = 10$ free variables and is much more tractable. This problem can be solved by any reasonable nonlinear least squares optimizer; we used MATLAB's \texttt{lsqnonlin} function. We chose $N = 6$ images from an eclipse observed by SWAP on 2011~June~1 and obtained the PSF shown in Figures~\ref{fig:psf_profile} and \ref{fig:2d_psf}.

\begin{singlespace}
  \bibliographystyle{apj} 

\end{singlespace}

\end{document}